\documentclass[twocolumn,%showpacs,
prl
,superscriptaddress
]{revtex4}

\usepackage{subfloat,float}
\usepackage[caption=false]{subfig}
\usepackage{amssymb}
\usepackage{amsmath}
\usepackage[pdftex]{graphicx}
\usepackage{epstopdf}
\usepackage{color}
\usepackage{comment}

\begin{document}
\author{Marianne Bauer}
\email{Marianne.Bauer@physik.uni-muenchen.de}
\author{Erwin Frey}
\email{frey@lmu.de}
\affiliation{%
Arnold-Sommerfeld-Center for Theoretical Physics and Center for NanoScience, Theresienstr. 37\\
Department of Physics, Ludwig-Maximilians-Universit\"at M\"unchen, D--80333 Munich, Germany}

\date{\today}

\title{On the role of multiple scales in metapopulations of public good producers}
\begin{abstract}
	Multiple scales in metapopulations can give rise to paradoxical behaviour: in a conceptual model for a public goods game, the species associated with a fitness cost due to the public good production can be stabilised in the well-mixed limit due to the mere existence of these scales.
	The scales in this model involve a length scale corresponding to separate patches, coupled by mobility, and separate time scales for reproduction and interaction with a local environment. 
	Contrary to the well-mixed high mobility limit, we find that for low mobilities, the interaction rate progressively stabilises this species due to stochastic effects, and that the formation of spatial patterns is not crucial for this stabilisation.  
  \end{abstract}
\maketitle

\begin{figure}
	\includegraphics[width=\columnwidth]{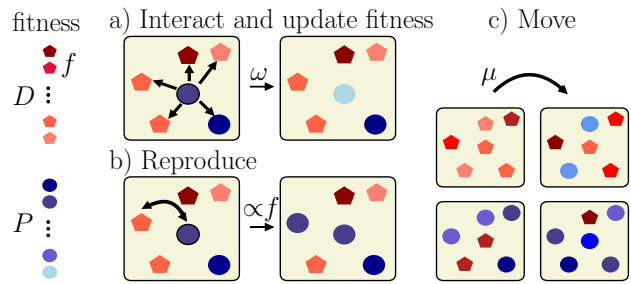}
	\caption{We study a public goods game between the good's producers (blue circles) and non-producers/defectors (red pentagons) with separate interaction (fitness update) and reproduction steps, which both take place locally on a patch $\nu$. The fitness is updated when bacteria sense their local environment with rate $\omega$ (a). Reproduction occurs via a Moran process  proportional to a player's fitness $f$ (b).  The separation of these steps means that players with different fitness values are present (colour shades). We study patches on a two-dimensional lattice, where a hopping rate $\mu$ couples different patches (c).\label{fig:model}}
\end{figure}

Ecological populations have evolved into systems where a variety of length and time scales exist due to patterns, heterogeneities, or temporal variations in the states of individual agents~\cite{Levin2000}. For example, populations dwell in spatial sub-structures of separate or fragmented habitats, in which the species form a tight community and are essentially well-mixed \cite{Hanski98,Black2012}. On larger scales, these habitats are then coupled by migration or dispersal. Similarly, individuals take up nutrients and interact with their local environments more frequently than they reproduce or die. The impact of such multiple length and time scales on ecosystems is associated with the maintenance of biodiversity, where separate scales can stabilise subpopulations which would otherwise die out~\cite{DurrettLevin, KorolevRevModPhys}. 

In order to investigate this complex interplay of different scales, we study a simplified model system for two different species with interacting individuals. With this model, we can tune the frequency of interaction events and dispersal, and thus evaluate the importance of the coupling between two specific length and time scales. We show that even for very simple interactions, the interplay of these scales leads to counterintuitive phenomenology. % due to a cross-over from stochastic to deterministic dynamics. 
Indeed, recent advances in both microbial~\cite{CelikerGore, GandhiKorolevGore} as well as interacting systems built with nucleic acids~\cite{Baccouche2014,ZadorinEstevezTorres,UrtelBraun} mean that such scales will be accessible experimentally, which helps elucidate ecological principles and control matter on small scales, respectively~\cite{andrianantoandro,Jessup2004,Padirac2013}. Theoretical models can assist both these experimental advances by identifying what interaction principles are ecologically viable or can lead to interesting effects in these experiments.
%Conceptual theoretical models can assist both these experimental advances by providing a framework for the type of interactions and spatial setups that can lead to interesting phenomenological behaviour.

A well-understood type of interaction are public goods games, which have been demonstrated in a variety of microorganisms, from bakterioplankton to yeast~\cite{diggle07, WestGriffinGardnerDiggle,  GreigTravisano, GoreOudenaarden,CorderoPolz12,Drescher2014,Becker17}. 
In such games, players receive fitness benefits due to the presence of a public good, which is produced only by a part of the population (`producers')~\cite{Velicer2003}.
In theoretical models for this scenario~\cite{MaynardSmith,HofbauerSigmund,frey2010}, such as the prisoner's dilemma~\cite{Hardin1968,Axelrod1981}, public good production incurs a fitness cost for the producers.  Due to this cost, producers in well-mixed populations in the prisoner's dilemma always die out.  
Thus, we will analyse the impact of the interplay of time scales (associated with the frequency of interaction events) and length scales (associated with spatial structure) for this game as deviations from this simple outcome.
%Since deviations from this outcome are easy to analyse, we investigate the impact of the interplay of time and length scales for this game.

Here, we discuss how reducing fitness updates (associated with an interaction or sensing rate) progressively stabilises producers up to finite costs in the small mobility limit. In the large mobility limit, the mere existence of %separate time scales for reproduction and sensing 
multiple scales (temporal interaction frequency, and spatial structure)
leads to a sharp jump in the stability of producers, as compared to a situation where reproduction and interaction occur in one step. Thus, producers are paradoxically stabilised by increasing mobility in this system, even though in diffusive well-mixed systems producers should die out~\cite{enquistleimar,dugatkin, Hamilton64}.

%%%%%%

\section{Model}
We consider a metapopulation of spatially separate subpopulations on patches arranged on a square lattice (Fig.~\ref{fig:model}). Patches initially contain $N_0$ players each, and are coupled by hopping with rate $\mu$ (Fig.~\ref{fig:model}~c), which then changes the number of players on a patch. Interactions only happen locally on a patch $\nu$ of this metapopulation~\cite{LevinPaine74, Chesson85, Tilman94, NeeMay94, Hanski98}, as opposed to between players on different patches~\cite{Nowak92,Szabo2007, RocaSanchez09,PercSzolnoki2010, SantosPacheco}.

The  interaction rate $\omega$ determines how often a player interacts with its environment locally on its patch and then updates its fitness $f$ to reflect the environment accordingly (see Fig.~\ref{fig:model}~a).
As players retain their fitness from the time when they last  sensed their environment,
 the fitness of players on a patch can vary (colour shades in Fig.~\ref{fig:model}). 
In microbes, the fitness generally increases with the amount public good on a patch, which in turn increases with the number of producers $P$ on a patch $\nu$ of $N_\nu$ players. Microbial experiments have shown that the relationship between the fitness and the numbers of producers on a patch  can be complex, and often nonlinear~\cite{DamoreGore, GoreOudenaarden}.
Since we are interested in a conceptual model that may apply to both ecological as well as engineered populations, we opted for a simple and general form for the fitness. Thus, we assume that the fitness obtained during such an interaction event scales linearly with the number of producers $P$ on a patch of $N_\nu$ players (see e.~g.~\cite{MitteldorfWilson, Taylor92, vanBaalenRand, DoebeliHauert, ArchettiScheuring, CremerMelbingerFrey}). We note that in these models, producers can also profit from the public good that they produce themselves.
The assumption that the fitness depends linearly on $P$ means that our model is conceptual and amenable to simple analysis. A different type of monotonously increasing fitness functions will not change our results for small mobility qualitatetively, and our arguments for large mobility can also be adapted to such nonlinear fitness functions.
  The fitness of the part of the population that does not produce the public good (`defectors' $D$) on a patch with $N_P$ producers is thus
\begin{equation} \label{eq:1} f_{N_P}^{D} = f_0+\frac{N_P}{N_\nu} b \textrm{,}\end{equation}
	where we set both the base fitness $f_0= 1$, as well as the benefit of the public good $b=1$, to one. The fitness of a producer on a patch with $N_P$ producers is lower than that fitness by a cost $c$, which corresponds to the cost of producing the public good: \begin{equation} f_{N_P}^{P}=f_{N_P}^{D}-c \textrm{.} \end{equation}
 We consider $c<1$ so that the benefit is higher than the cost of public good production.

We assume that players on a given patch reproduce proportionally to their fitness according to the frequency-dependent Moran process~\cite{Moran62}, by replacing another randomly selected player on the same patch (Fig.~\ref{fig:model}~b)~\cite{TaylorNowak, Traulsen05, Traulsen06, traulsen2009stochastic,Gelimson, Ashcroft17}. In order to assure that players cannot reproduce themselves, we defined reproduction rates as $r=f (1-1/N_\nu)$, such that in this work, the largest possible reproduction rate for a producer at zero cost for the mean number of players per site is $r_P= 2(1-1/N_0)$, and the minimal reproduction rate for a defector is $r_D= (1-1/N_0)$~\footnote{This rescaling of our reproduction rates ensures that a single player on a site can never reproduce at all. We verified that our results are not affected by this rescaling factor in way other than its scaling by strictly setting reproduction rates for a single player per site to zero in some simulation runs.}.
 
We focus on $\omega>r_P$ throughout, such that players have on average updated their fitness before reproducing.

Initially, equal numbers of defectors and producers are distributed randomly in the metapopulation of $L\times L$ patches.  %We set $N_\nu=6$ and $L=20$ unless otherwise specified. This $L$ is large enough for all results shown to be independent of $L$, which we discuss later.
We use a Gillespie algorithm~\cite{Gillespie} to simulate the stochastic dynamics of the system  until either producers or defectors have gone extinct.

\section{Low mobility limit}
We will start by elucidating the dynamics of the metapopulation in the small mobility limit, $\mu \ll r_P < \omega$. Since hopping is rare in this limit, patches on the metapopulation fixate to either producers or defectors before the first movement occurs.
When the number of individuals on each patch is small, $N_0 c \ll 1$, the dynamics is dominated by demographic fluctuations while fitness differences play only a subordinate role \cite{Cremer2009}. %Since we study a large field of patches such that individual fluctuations do not matter, and 
Indeed, in our system we are mostly concerned with values of cost $c \ll 1$ (as we will see in the following), and so this criterion applies. Thus, the number of patches on which producers fixate is approximately equal to the number of patches where defectors fixate.

On timescales longer than fixation, hopping events occur, whereby a single player hops from its patch to a neighbouring patch. For small mobility, this new patch will likely fixate again before another player hops. The process of fixation on the entire metapopulation can thus only occur via a series of single player hops and subsequent fixation on single patches.
One can naively assume that if the probability that a producer invades a patch of defectors is higher than the probability that a defector will invade a patch of producers, producers will fixate in the metapopulation. 
 Hence, we will first concentrate on the probability that a player successfully invades a patch fixated to the other type. 

\begin{figure}
        \includegraphics[width=0.5\textwidth]{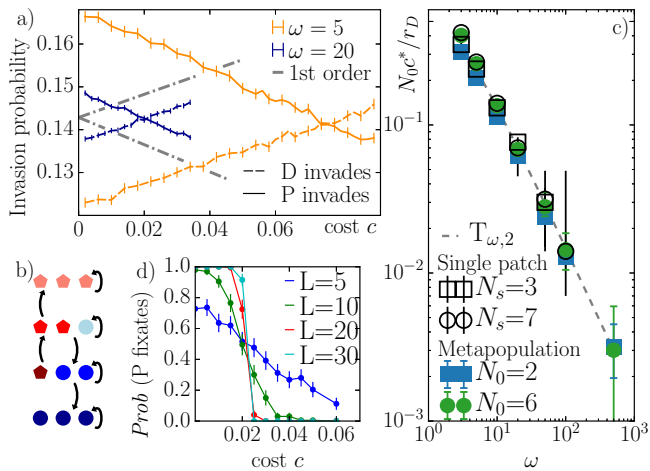}
	\caption{Cost $c^{*}$ up to which producers fixate for small mobilities: a) Single patch invasion probability of  a producer (solid) or defector (dashed lines) on a patch of $N_\nu=N_0=6$ defectors or producers, respectively (i.e. seven players total, $N_s=7$).
	Smaller $\omega$ (error bars show 95\% confidence interval) increases the cost of equal invasion probability $c^{*}$, from the $\omega \to \infty$ limit. 
	b) Possible transitions (arrows) between player configurations in the transition probability matrix. c) Rescaled $c^{*}$  for a variety of $N_0$ in a metapopulation ($L=20$) (filled symbols) agrees with the corresponding $c^{*}$ from single patch invasions (empty symbols), and scales linearly in $1/\omega$. The estimate from $T_{\omega,2}$ for $N_0=2$ players  correctly shows this scaling with $\omega$. d) Fixation probability $P$ for $\omega=20$ in the metapopulation becomes sharper, but does not change its value for increasing $L$. \label{f:inv}}
\end{figure}

\subsection{Single patch invasion}
Players on fixated patches all have the same fitness (fitness $f_{N_0}^{P}=1+  N_0/N_0-c = 2-c$ on patches with producers, and the lowest possible fitness of $1$ for defectors). Immediately after a hopping event, the invading player will retain its fitness from its previous (fixated) patch if the interaction rate $\omega$ is finite. %, meaning that it also has fitness $2-c$ ($1$) if it is a producer (defector).
Thus, an invading producer has an advantage on the stack of defectors initially, due to its higher fitness, while the opposite is the case for the invading defector.
This advantage of the producer occurs for a variety of models with delayed fitness updates~\cite{Roca2006,Huang,YiZuwang,Miekisz} or memory~\cite{kirchkamp}. Here, we investigate its impact by tuning the rate $\omega$. In the following, we show how one can understand the impact of this rate $\omega$ on fixation using Markov chains, independently of the precise choice of the fitness function.

In order to emphasise that we are discussing a single patch in the following, we refer to the number of players on this single patch by $N_s$. Since we are concerned with cases where one player of one type fixates on a patch with $N_0$ players of the other type, the $N_s$ relevant for understanding the metapopulation is $N_s = N_0 +1$. We refer to this successful single patch fixation probability of one player of a particular type as the invasion probability.

\subsection{Immediate interaction with the local environment}

We start with the invasion probability of a producer (defector) on a patch of defectors (producers) in the limit of $\omega \to \infty$, before decreasing $\omega$.
Then, interactions occur so often that the fitness values of players always reflect their local environment, and, in particular, that the fitness of the invading player is updated immediately.

\emph{Transition probability matrix.}
The probability that this invading player fixates can be calculated by considering the transition probability matrix for the corresponding Markov chain~\cite{KarlinTaylor, BlytheMcKane}. This transition probability matrix is constructed by calculating the transition probabilities from any possible state, corresponding to any possible configuration of players on this patch, into any other possible state.
For $\omega \to \infty$, the fitness of all $N_s$ players is always updated to reflect the configuration on the patch. Since all fitness values are then uniquely determined by the numbers of producers on a patch (see Eq.~\ref{eq:1}), there are exactly as many states as there are configurations of different numbers of producers on the patch. In general, there are $N_s+1$ such states (for zero up to $N_s$ producers on the patch).  For $N_s=3$, we sketch these states in Fig.~\ref{f:inv}~b, with the different fitness values encoded in the colour scale. The arrows indicate which states a particular state can transition into through reproduction of a player. The corresponding transition probabilities are proportional to the reproduction rate of that player.

	The probabilities to be in one of these different configurations make up the $N_{s}+1$- dimensional probability vector. More precisely, the $i^{\textrm{th}}$ entry in our probability vector corresponds to the configuration with $i$ producers, for $i\in \{0,N_{s}\}$. The fitness of defectors in state $i$ is then $f_{i}^{D}=f_0+i/N_{s}=1$ and $f_{i}^{P}= f_{i}^{D}-c$. Thus, the transition probability matrix for $N_{s}=3$ reads 
\begin{align*} \label{e:3}
T_{3}=
\left( {\begin{array}{cccc}
	1 & T_{1}^{-} &0 &0\\
	0 & T_{1}^{d} & T_{2}^{-} & 0\\
	0 & T_{1}^{+}  & T_{2}^{d} & 0\\
	0 & 0           & T_{2}^{+} & 1\\
\end{array} } \right) \textrm{,}
\end{align*}

	where $T^{-}_{i}$ denotes the probability to transition from state $i$ to state $i-1$, and $T^{+}_{i}$ the probability to transition from state $i$ to $i+1$. In our case, $T_{1}^{-}=f^{D}_{1}/(2f^{D}_{1}+f^{P}_{1})$ corresponds to the probability that a defector in a state with one producer replaces that producer, while $T_{1}^{+}=f^{P}_{1}/(2f^{D}_{1}+f^{P}_{1})$ denotes the probability that the producer replaces one defector. Similarly,  $T_{2}^{-}=f^{D}_{2}/(f^{D}_{1}+2f^{P}_{1})$ and $T_{2}^{+}=f^{P}_{2}/(f^{D}_{1}+2f^{P}_{1})$ correspond to analogous events for the state with two producers.
The diagonal elements $T_{1}^{d} = 1-(T_{1}^{-}+T_{1}^{+})$ ($T_{2}^{d}=1-(T_{2}^{-}+T_{2}^{+})$) contain the probabilty that a defector (producer) replaces another player of its type. Since this is a stochastic probability matrix, the entries of a columns add up to one.

	Repeated iteration of this transition probability matrix on the initial state (in our case, that containing only one producer) converges to a probability distribution with two non-zero elements corresponding to the absorbing states~\cite{KarlinTaylor} (in our case, the first and last elements, corresponding to states with only defectors or only producers). The transition probability from a state of one producer into the absorbing state with all producers (which we refer to as invasion probability $I$) is known analytically from a first step analysis~\cite{VanKampen,AntalScheuring, AltrockTraulsen,KarlinTaylor}, and reads
\begin{equation}
	I_P=\frac{1}{1+\sum_{i=1}^{N_{s}-1} q_i}\textrm{,}
\end{equation}
where $q_i=\prod_{j=1}^{k} \frac{T_{j}^{-}}{T_{j}^{+}}$.

\emph{Invasion probability to first order linear in $c$.}
For small cost, an expansion of this equation in cost is accurate and more intuitive. To first order in $c$, the probability that a single producer successfully invades a patch of defectors for the $\omega \to \infty$ limit reads  
\begin{equation}
	I_P\approx\frac{1}{N_s} - \frac{c}{N_{s}^2} \sum_{i=1}^{N_s-1} \frac{N_s - i}{f^{D}_{i}} \textrm{,} 
\end{equation} where $f^{D}_{i}$ is the fitness of a defector in a state with $i$ producers. Intuitively, the sum is over all states that need to be transitioned through in order to arrive at the absorbing state; each term contains the number of defectors that can be replaced in that state, and their fitness. The analogous expression for invasion probability of a defector is  
\begin{equation}I_D \approx \frac{1}{N_s} + \frac{c}{N_{s}^2} \sum_{i=1}^{N_s-1} \frac{i}{f^{D}_{i}}
\end{equation}. This expansion only depends on the fitness being linear in $c$, not on the linearity of the fitness function with respect to public good producers (provided that the states with only producers and defectors are the two absorbing states).

Figure~\ref{f:inv}~a shows the invasion probability from this expansion (grey dashed-dotted line) for $N_s=7$. The solid decreasing line (dashed increasing line) corresponds to the invasion probability for a producer (defector). The invasion probabilities are equal only at zero cost, where they have the value  $1/N_s$, i.e. the probability for fixation of one out of $N_s$ players in a Moran process. As the invasion probability of producers decreases with increasing $c$, this implies that for $\omega \to \infty$, producers die out.

\subsection{Delayed interaction with the environment}

Using simple arguments from Markov chains, we have learnt that the probability for successful invasion (or fixation) of a producer into a patch of defectors tends to be lower than the invasion probability of a single defector into a patch of producers, apart from at zero cost, where the two are equal. Now, we ask how a delay in interactions (and thereby in adjusting fitness to reflect the current environment) affects these invasion probabilities by considering finite interaction rates $\omega$.

\emph{Transition probability matrix: increase in state space}
For finite $\omega$, additional states corresponding to players with different fitnesses from previous configurations would need to be included. 
These additional states correspond to configurations where the fitness of some players is not updated, but corresponds to their previous environment. For simplicity, we explain the structure of the transition probability matrix for $N_s=2$ players with one invading players, i.e. $N_s+1=3$ players per patch, in the appendix. We note here that a brute force first step analysis (or diagonalisation) of these transition probability matrices becomes unintuitive already for such a small number of players. Numerical solution is possible, and indeed, we discuss and show such a numerical result later for comparison. For now, we turn to numerical simulations of the invasion probabilities on a single patch.

 We show these numerical simulations for invasion probabilities for $N_s=7$ for both with $\omega =20$ and $\omega=5$ in Fig.~\ref{f:inv}~a, where error bars denote $95\%$ confidence intervals. We note that these lines are approximately parallel to the first order expansion for $\omega \to \infty$, meaning that finite $\omega$ leads to a parallel shift in invasion probabilities. 
 
 Thus, we find that the effect of reducing the frequency of interactions is that the invasion probabilities for producers and defectors are shifted up- or downwards, respectively. This shift means that for decreasing $\omega$, the invasion probabilities are equal at a finite cost $c^{*}$. 

\subsection{Cost $c^{*}$ at small mobility} After having understood the single patch invasion probabilities, we turn to the metapopulation of $L\times L$ patches. In Fig.~\ref{f:inv}~c, we show this $c^{*}$, the highest cost up to which producers can fixate, for a variety of $\omega$ and two different $N_0$. 

We take $c^{*}$ for the metapopulation to be the cost where exactly half the simulation runs fixate to producers. The transition from producers fixating in all simulations to defectors fixating in all simulations is increasingly sharp for larger  $L$ (Fig.~\ref{f:inv}~d for $\omega=20$, $\mu=0.05$ and 200 simulations). Thus, this $c^{*}$ is the highest cost up which producers can fixate in large systems. As its value does not change with $L$, it is sufficient to use $L=20$ for the values of $c^{*}$ presented in this work.
Error bars in Fig.~\ref{f:inv}~c  mark the costs where 30 \% or 70 \% of all runs fixate to producers.
These error bars are essentially negligable for our system sizes, even for a relatively small number of 30 simulation runs, except for large $\omega \gtrapprox 500$ (see Fig.~\ref{f:inv}~d).

Producer fixation for $N_0=6$ players per patch on the metapopulation is intended for comparison with the single patch invasion of $N_s=7$ just discussed (six players of one type per patch, being invaded by one player of the other type after a hopping event).

We first concentrate on comparing producer fixation in the metapopulation with what we would expect from the single patch invasion probability: The empty symbols in Fig.~\ref{f:inv}~c) show the cost at which single patch producer invasion is as likely as single patch defector invasion (for $N_s=3$ and $N_s=7$), while the filled symbols show the highest cost at which producers fixate in our metapopulation (for $N_0=2$ and $N_0=6$). We note that we need to compare metapopulations with $N_0$ players per patch with single patch invasions of $N_s=N_0+1$ players.
For $N_0=2$,  $c^{*}$ from the single patch invasion probabilities slightly overestimates $c^{*}$ obtained from the metapopulation. The single patch invasion argument overestimates in this case because the noise in the number of players on a patch in the metapopulation, caused by hopping, has a comparatively large effect for this small $N_0=2$. 
For $N_s=6$, single patch and metapopulation results for $c^{*}$ agree very well.

Indeed, the results from the metapopulation also agree with the numerical value of $c^{*}$ obtained from the single patch transition probability matrix $T_{\omega,2}$ for $N_0=2$ (dashed line), by repeated numerical iteration on the initial state as discussed in the appendix. Thus, the fact that single patch dynamics determine the result for  $c^{*}$ indicates that the question of which species fixates in the metapopulation is decided by invasion of single patches, rather than by spatial effects. Hence spatial effects or pattern formation - often the cause for stabilising producers in spatial systems - plays no role in determining the maximal cost for stabilising the producers in a system where interactions take place locally.

Before moving on to higher mobilities, we point out that the values for $c^{*}$ (rescaled by $N_0$) in Fig.~\ref{f:inv}~c scale linearly in $1/\omega$, independently of the number of player per patch $N_0$ studied here, up to numerical accuracy. Thus, at small mobilities $c^{*}$ scales inversely in interaction rate and number of players per patch.
%Due to the size of the state space (for $N_\nu=2$, this is a $14 \times 14$ matrix), we only calculated this matrix analytically for $N_\nu=2$ as a proof of principle (see supplement~\cite{ref:supp}). We presume that higher $N_\nu$ would improve the value of $c^{*}$ obtained.

\begin{figure}
	\includegraphics[width=0.5\textwidth]{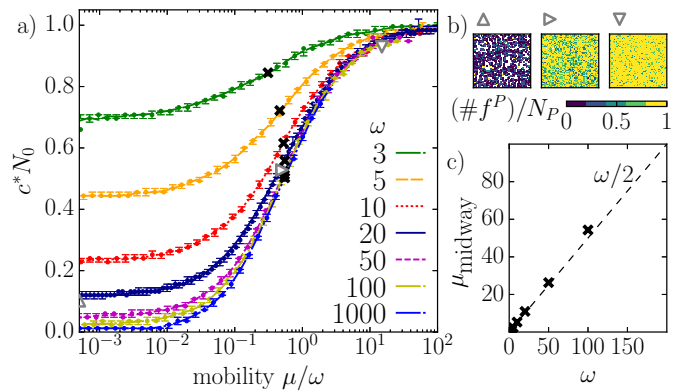}
	\caption{a) Cost $c^{*}$ (rescaled by $N_0=6$) for different interaction rates $\omega$ (lines serve as guide to the eye) and $L=20$: For large mobilities, producers are stabilised and $c^{*}=1/N_0$, as the fittest player in the metapopulation is a producer when production cost does not outweigh the contribution from a single player. %This $c^{*}$ is larger than $c^{*}= r_P/(\omega N_\nu)$ at small mobilities. 
	b) Snapshots of the number of different fitness values per producers on a patch ($L=50,\omega=20$), for parameters shown by triangles in a): at small $\mu$, patches are occupied by one type of player only. For higher $\mu$, producers spread across patches, and the number of different fitness types per patch increases. c)  $c^{*}$ in a) reaches its midpoint at $\mu\approx \omega$ (crosses). %, showing that the interplay of these two rates determines the outcome of the game.
	\label{fig:phase}}
\end{figure}

\section{Increasing and high mobility.}
We thus established that for low mobilities, the producers are stabilised by less frequent interactions up to a finite cost $c^{*}$, which scales as $1/(N_0 \omega)$. Figure~\ref{fig:phase}~a shows this $c^{*}$ for a variety of hopping rates $\mu$ and interaction rates $\omega$. For all $\omega$, $c^{*}$ saturates for low mobilities: as long as invasion is the dominating process, the value of the hopping rate is irrelevant. Paradoxically, $c^{*}$ increases, and thus production stabilises, upon increasing mobilities, where one would normally expect producers to die out~\cite{enquistleimar,dugatkin, Hamilton64}. Stabilisations of producers have been found in more complex models, for example with smarter strategies, or evolving mobility~\cite{Constable2016, VainsteinArenzon,Aktipis, MeloniMoreno, Ferriere, LeGalliard05}. Here, our simple model differs from a normal prisoner's dilemma only by the reduced interaction frequency and coupled patch geometry. It is thus interesting that we find stabilisation of producers in the high, as compared to the low, mobility limit. Next, we explain the stabilisation in the high mobility limit. %, by focusing on $c^{*}$ in a well-mixed system. 

For high mobilities, $r_P < \omega \ll \mu$, players can be considered as essentially well-mixed. This is the case as long as the number of patches $L \times L$ is large enough, so that mixing eliminates correlations, i.e. the total number of players in the metapopulation should be $N \gg N_0$.
Indeed, almost all players on a patch for $\mu=300$ (last snapshot in Fig.~\ref{fig:phase}~b) have different fitness values, and producers and defectors both occur on each patch.

Since $\mu \gg \omega$ in this limit, the precise value of the rate $\omega$ is irrelevant (as long as we maintain $\omega>r_P$, so that players have adjusted their fitness to their environment before reproduction), and all possible different fitness values of players will occur. There are as many possible different fitness values as there are different player configurations on a patch. In the high mobility limit,  the number of players on a patch can be different from $N_0$ due to hopping, and more than $N_0+1$ fitness values per species are possible in principle. However, since the number of players in the metapopulation is constant and the mean number of players per site is $N_{0}$, we can focus on the dominant $N_0+1$ fitness values. Thus, we can consider reproduction as a Moran process of all $N$  players in the well-mixed metapopulation. Since $N$ is large, we are in a deterministic limit, where the fittest species fixates in the entire metapopulation.  In order to determine if producers or defectors survive, we thus need to compare the fitness of the individual players, and will start with the fittest producer and defector.
%\ch{}{We will do so for a generalised version our linear fitness function, namely a fitness function of the type $f_0 + g(N_P/N_\nu)b$ for defectors, where $g$ is any monotonously increasing function.}

%Since $\mu \gg \omega$ in this limit, the precise value of the rate $\omega$ is irrelevant (as long as we maintain $\omega>r_P$, so that players have adjusted their fitness to their environment before reproduction), and all possible different fitness values of players will occur.
%There are as many possible different fitness values as there are different player configurations on a patch. In the high mobility limit,  the number of players on a patch can be different from $N_\nu$ due to hopping, and more than $N_\nu+1$ fitness values per species are possible in principle. Since the mean number of players per site is $N_{\nu}$, we can nevertheless focus on the dominant $N_\nu+1$ fitness values. Thus, we can consider reproduction as a Moran process of all $N$  players in the well-mixed metapopulation. Since $N$ is large, we are in a deterministic limit, where the fittest species fixates in the entire metapopulation.  In order to determine if producers or defectors survive, we thus need to compare the fitness of the individual players, and will start with the fittest producer and defector. 

The fittest producer is surrounded only by producers, and so its fitness is $1+ N_0/N_0-c =2-c$. The fittest defector is also surrounded only by producers, but since it is not a producer itself, its environment contains $N_0-1$ producers; thus, its fitness is $1+ (N_0-1)/N_0 =2-1/N_0$. Similarly, all other defector types have fitnesses lower by $1/N_0$ than the corresponding producer fitness type at zero cost. Thus, if $c\le 1/N_0$, all producer fitness values are equal or higher to the corresponding defector fitness values. Producers will thus fixate for $c\le 1/N_0$, which corresponds exactly to the high mobility limit in Fig.~\ref{fig:phase}.  This value is independent of $\omega$, and thus the mere existence of the patch structure and separate time scale for interaction and reproduction stabilises producers in this limit.

 The term  $1/N_0$ measures the impact of one single player on the fitness of players (self-interaction). The fact that this stabilisation occurs in the high mobility limit is intriguing: it is reminiscent of Hamilton's rule of relatedness~\cite{Hamilton64}, where production can be stable for \emph{small} dispersal speeds in viscous populations~\cite{MitteldorfWilson, Taylor92, vanBaalenRand} or on graphs and social networks~\cite{OhtsukiHauertNowak}, or of group selection that can occur in structured populations. Here, the stabilisation arises in the \emph{high} mobility limit for a different reason: fit producers exist for all mobilities, as time scale separation and patch structure mean that players only sense the fitness within small subgroups. The high mobility allows these fit producers to take over the population, similarly to how fit species dominate adapting populations \cite{NeherHallatschek}.

%We have now understood the limiting values for $c^{*}$  for low and high mobilities in Fig.~\ref{fig:phase}~a.
%For all interaction rates, $c^{*}$ increases smoothly from the low mobility limit at  $c^{*}\approx r_P /(N_\nu \omega)$ to the high mobility limit at $c^{*}=1/N_\nu$.
 In  Fig.~\ref{fig:phase}~a, $c^{*}$ increases smoothly from the low mobility limit at  $c^{*}\approx r_P /(N_0 \omega)$ to the high mobility limit at $c^{*}=1/N_0$ for all interaction rates.
We note that for high $\omega$, $c^{*}$ increases more slowly than for low interaction rates (logarithmic x-axis). Indeed, Fig.~\ref{fig:phase}~c shows that $\mu$ at which $c^{*}$ reaches its midpoint is proportional to $\omega$. Thus, the interplay between these two rates ($\mu$ for the spatial separation and $\omega$ for the interaction) selects when stochastic invasion probabilities and when the fitness of the fittest player determine the survival of producers.

\section{Summary and Outlook}
In the framework of a  prisoner's dilemma, the coupling of spatial length scales and temporal separation of interaction and reproduction can lead to  stabilisation of players whose fitness is associated with a cost. 
We have shown here that this stabilisation manifests itself differently for low and high mobilities. For low mobilities, the stabilisation is proportional to increasing separation of interaction and reproduction; for high mobilities, the stabilisation up to $c^{*}=1/N_0$ occurs as long as these time scales are separate and as long as there is spatial structure. We also found that the formation of spatial structure or patterns plays no role in stabilisation of producers in a setup where interactions and reproduction take place in a well-mixed local environment. 

Producers are also stabilised to higher costs for high than for small mobilities in other public goods games, such as the snowdrift game, if self-interaction is included.
%Similar reasoning applies to other public goods games: for the snowdrift game, where producers and defectors can coexist for small costs, producers are also stabilised to higher costs for high than for small mobilities, if self-interaction is included.
As the interplay of length and time scales can give rise to such counterintuitive results in simple models, it deserves further attention in order to guide what models may be worth exploring with synthetic biological systems. 

\section{Acknowledgement}
We are grateful to S. Rulands for discussions and insights into evolutionary games, and N. Goldenfeld for helpful discussions and references. MB thanks S. K. Baur for multiple discussions concerning transition probability matrices, and I. Graf, M. Rank and C. Weig for feedback. We acknowledge funding from a Marie-Sk\l{}odowska Curie Horizon 2020 grant, and the German Excellence Initiative via the programme `NanoSystems Initiative Munich'.

\section{Appendix}
In this appendix, we show and explain the transition probability matrix with  $N_0=2$ players with one invading players, i.e. $N_0+1=3$ players per patch. For every state with $N_P \in \{0,N_0 \}$ producers there are now additional states with either one, two or all three players with fitnesses from their previous patch, corresponding to five additional states.  Since it does not matter whether the absorbing states are updated or not, we can ignore the additional absorbing states (with all producers or defectors, but varying number of players with updated fitness). Thus, we obtain a total of fourteen states. 

These states are structured as follows: states 1-4 are shown in Fig.~\ref{f:inv}~b. State 5-14 are structured such that all states with odd numbers contain two producers and one defector, and all states with even numbers contain two defectors and one producers. In states 5 and 6, a single producer retains fitness $f^P$ from it's previous patch, and in states 7 and 8, a single defector retains its fitness $f^D=1$ from its previous patch. States 9-12 contain two players that retain their fitness: in states 9 and 10, only one producer is fully updated, and in states 11-12, only a single defector is fully updated. In states 13 and 14, all players retain their fitness from their previous patch: in state 13, the two producers have fitness $f^P=2-c$ and the defector has fitness $f^D=1$, while in state 14, the producer has fitness $f^P$ and the two defectors have fitness $f^D$. 
Our reproduction rate  $r^{D/P}_i$ is proportional to the corresponding fitness. %~\footnote{For most of this work, the reproduction rates $r^{(P,D)}_{i}$ are related to the fitness values by $r^{(P,D)}_{i}=(1-1/3) f^{(P,D)}_{i}$, with $f^{D}_1=1+1/3$, $f^{D}_2 = 1+2/3$ and analogously for producers, which ensures that single players on a patch cannot reproduce}. 
 The transition matrix thus reads

\begin{widetext}
	\tiny{
\begin{align*}
	&T_{\omega,2}=\\&
	\left( 
\arraycolsep=1.4pt\def\arraystretch{2.2}
	{\begin{array}{cccccccccccccccc}
		1 & r^{D}_{1}/F_1&0            & 0 & 0                             & r^{D}_{1}/F_6          & (r^{D}{+}r^{D}_{1})/F_7   &  0             & 0                        & r^{D}/F_{10}            &0         & (r^{D}{+}r^{P}_{1})/F_{12} & 0 & 2r^{D}/F_{14}\\
		0 & r^{D}_{1}/F_1& r^{D}_{2}/F_2 & 0 & r^{D}_{2}/F_5               & 2\omega/F_6            & (r^{D}_{1}{+}2\omega)/F_7 &  0             & 0                        &0                        &0         & 0 & 0 & 0 \\
		0 & r^{P}_{1}/F_1& r^{P}_{2}/F_2 & 0 & (r^{P}_{2} {{+}}2\omega)/F_5& 0                      & r^{P}_{1}/F_7        &  2\omega/F_8        & 0                        &0                        &0         & 0 & 0 & 0 \\
		0 & 0          & r^{P}_{2}/F_2 & 1 & (r^{P}_{2} {+}r^{P})/F_5      & 0                      & 0                    &  2r^{P}_{2}/F_8     &(r^P{+}r^{P}_{2})/F_9     &0                        &2r^{P}/F_{11} & 0 &2r^{P}/F_{13}& 0 \\
		0 & 0          & 0           & 0 & 4\omega/F_5                    & 0                      & 0                    &  0                  &2\omega/F_9               &0                        &4\omega/F_{11}& 0 & 0 & 0\\
		0 & 0          & 0           & 0 & r^{D}_{2}/F_5                   &(2r^{D}_{1}{+}4\omega)/F_6& 0                 &  0                  & 0                        & 0                       &2 r^{D}_{2}/F_{11} &( r^{D}_{1}{+}2\omega)/F_{12} & 0 & 0 \\
		0 & 0          & 0           & 0 & 0                               & 0                      & 4\omega/F_7         &  0                  & 0                        & (2r^{D}{+}2\omega)/F_{10}&        0 & 2\omega/F_{12} & 0 & 0 \\
		0 & 0          & 0           & 0 & 0                               & 0                      & r^{P}_{1}/F_7        &(2r^{P}_{2}{+}4\omega)/F_8&(r^{P}_{2}{+}2\omega)/F_9& 2r^{P}_{1}/F_{10}   &  0  & 0 & 0 & 0 \\
		0 & 0          & 0           & 0 & 0                               & 0                      & 0                    & 0                   & 2\omega/F_9              & 0                        &       0  & 0 & 4\omega/F_{13}& 0 \\
		0 & 0          & 0           & 0 & 0                               & 0                      & r^{D}/r_7            & 2r^{D}/F_8          & r^{D}/F_9                &(r^{D}{+} 2\omega)/F_{10}  & 0        & 0 & 0 &2\omega/F_{14} \\
		0 & 0          & 0           & 0 & r^{P}/F_5                       &2r^{P}/F_6              & 0                    & 0                   &  0                       &  0                       &(2r_{C}{+}2\omega)/F_{11}& r^{P} /F_{12} & 2\omega/F_{13} &0 \\
		0 & 0          & 0           & 0 & 0                               & 0                      & 9                    & 0                   & 0                        &  0                       &0          & 2\omega/F_{12} & 0 & 4\omega/F_{14}\\
		0 & 0          & 0           & 0 & 0                               & 0                      & 9                    & 0                   & r^{P}/F_9                & 0                        & 0         & r^{P}/F_{12} &2 r^{P}/F_{13} & 2r^{P}/F_{14}\\
		0 & 0          & 0           & 0 & 0                               & 0                      & 9                    & 0                   & r^{D}/F_9                & 0                        & 0         & r^{D}/F_{12} & 2 r^{D}/F_{13} & 2 r^{D}/F_{14} \\
\end{array} } \right)  
	\textrm{,}
		\end{align*}
	}
\end{widetext}

where
$F_1= 2 r^{D}_{1} + r^{P}_{1}$, \\
$F_2=  r^{D}_{2} + 2r^{P}_{1}$, \\
$F_5 = 6 \omega + 2(r^{P} + r^{D}_{2} + r^{P}_{2})$,\\
$F_6 = 6 \omega + 2(r^{P} + 2 r^{D}_{1})$,\\
$F_7 = 6 \omega + 2(r^{D} + r^{D}_{1} + r^{P}_{1})$,\\
$F_8 = 6 \omega + 2(r^{D} +2 r^{P}_{2})$,\\
$F_9 = 6 \omega + 2(r^{D} + r^{P} + r^{P}_{2})$,\\
$F_{10} = 6 \omega + 2(2r^{D} + r^{P}_{1})$,\\
$F_{11} = 6 \omega + 2(2r^{P} + r^{D}_{2})$,\\
$F_{12} = 6 \omega + 2(r^{D} + r^{P} + 2r^{D}_{1})$,\\
$F_{13} = 6 \omega + 2(2r^{P}+ r^{D})$,\\
and $F_{14} = 6 \omega + 2(r^{P} + 2r^{D})$~\footnote{Since there are $N_0-1$ different reproduction events possible for each player (replacing every other player), the contribution for fitness update events also need to be multiplied by this number (i.e. doubled in this case).}.

	The successful invasion probability can be obtained by numerical iteration of this matrix on the probability vector corresponding to our initial state (corresponding to state 13 or 14 for an invading defector or producer, respectively), or alternatively by matrix inversion as detailed in Ref.~\cite{GrinsteadSnell}. 

%\bibliography{pd2}

\end{document}